\documentstyle[fleqn,twoside,gc,epsf]{article}

\heads{Bajkova}
      {Maximum Entropy Method for Reconstruction of the CMB Images}
\begin{document}
\twocolumn[
\Arthead{8}{2002}{pages (5)}{0}{0}

\Title{MAXIMUM ENTROPY METHOD FOR \yy
       RECONSTRUCTION OF THE CMB IMAGES}

   \Author{A.T.Bajkova\foom 1          
            }     
           {IAA RAS, Nab.Kutuzova, 10,
           191,187, St.Petersburg}      

\Abstract
    {We propose a new approach for the accurate reconstruction of cosmic
    microwave background distributions from observations containing in addition
    to the primary fluctuations the radiation from unresolved extragalactic
    point sources and pixel noise. The approach uses some effective
    realizations of the well-known maximum entropy method and principally
    takes into account {\it a priori} information about finiteness and
    spherical symmetry of the power spectrum of the CMB satisfying the Gaussian
statistics.}

] 
\email 1 {bajkova@quasar.ipa.nw.ru}

\section{Introduction}
Future cosmic microwave background (CMB) experiments, such as the PLANCK mission,
are intended to produce the hi-precision determination of the cosmological
parameters. Therefore it is necessary to ensure the adequate accuracy of
removing the foreground contamination from the galactic dust, free-free and synchrotron
emission, kinetic and thermal SZ effects from the galaxy clusters, extragalactic
point sources (PS) and pixel noise. All the foregrounds can be divided in two
main classes. The first one includes the so called low-multipole foregrounds
(from galaxy and galaxy clusters) which give significant contribution to the
total CMB observed signal in low spatial frequencies. The second one
represents the hi-multipole foregrounds from unresolved point sources and pixel
noise. In this paper we stress attention on the problem of separation of
the hi-multipole foregrounds. The problem of extraction of point sources is
considered by many authors. Tegmark et al\cite{3}, have proposed to use optimal
hi-pass filtering to suppress fluctuations due to CMB and to strengthen
the point sources with subsequent discarding from the input CMB map the pixels
contaminated by the detected point sources within a beam size. This method is
based on the idea that beginning from some value of multipole
$l_d$ (damping scale of the CMB fluctuations) the CMB signal is strongly
dominated by the point sources. The spectrum components between $l_d$ and and
$l_f$ ($l_f$ is a cut-off frequency determined by the antenna filtering
function), are considered as initial data for a chosen method of separation of
the point sources. In this paper we exploit the same idea and propose for
selection of the PS foreground to use the maximum entropy method (MEM) as a
very powerful technique for the image reconstruction from incomplete and noisy spectrum data.

The whole separation algorithm consists of the following four steps:

--deconvolution of the detected CMB map by the antenna beam profile to
obtain a (CMB+PS) estimation;

--MEM reconstruction of the PS foreground using those Fourier data of the (CMB+PS)
estimated map which are located in $(l_d, l_f$) multipole region;

--discarding from the estimated (CMB+PS) map those pixels which are
contaminated by detected PS and obtaining that a CMB estimation with the
missing components ('holes');

--MEM reconstruction of the missing CMB components using {\it a priori} information
about the finiteness and spherical symmetry of the CMB power spectrum.
(Note that the finiteness of the spectrum indicates the analyticity property of
CMB fluctuations. Therefore, in principle, the whole CMB signal can be
reconstructed from only a known part.)

It is expected, that the proposed approach is much more accurate than that
described one in Tegmark et al\cite{3}, because it allows to 1) detect the
point sources with much smaller amplitudes, and 2) reconstruct the discarded
CMB components. It necessary to note also that the reconstruction of the CMB
distribution in 'holes' allows to reconstruct the Gaussian nature of the
primary CMB fluctuations.

\section{The maximum entropy method}

In our simulations we use the MEM technique for the realization of three main
operations: 1) deconvolution of the input CMB map by antenna beam profile,
2) reconstruction of the PS foreground using the hi-multipole spectrum components,
3) reconstruction of the discarded CMB components. For solving the second task we use
a well-known standard maximum entropy method suitable for reconstruction of only
positive-definite signals. But for solving the first and third tasks a more
general version of the MEM (GMEM) (Bajkova\cite{1}) is needed. This version
enables to reconstruct the signals taking both positive and negative values.
First we consider the standard MEM approach.

\subsection{The standard maximum entropy method}

Let $x_{ml}, m,l=0,..., M-1$ be
unknown two - dimensional signal determined in $M$x$M$ discrete space. Then the
maximum entropy solution for $x_{ml}$ satisfies

\begin{equation}
-\sum_m\sum_l x_{ml}\ln x_{ml}\equiv H(x)=\max, ~x_{ml}\ge 0
\end{equation}
subject to constraints on $x_{ml}$ provided by data.

In all our tasks we deal with the linear constraints derived from
the Fourier-space data which may be represented as follows

\begin{equation}
\sum_m\sum_l x_{ml}\cos(2\pi(nm+kl)/M)+\eta_{nk}^r=D_{nk}^r,
\end{equation}

\begin{equation}
\sum_m\sum_l x_{ml}\sin(2\pi(nm+kl)/M)+\eta_{nk}^i=D_{nk}^i,
\end{equation}
where $D_{nk}^r, D_{nk}^i, 0\le n,k \le M-1 $ are data with superscripts
$r$ and $i$ denoting the real and imaginary parts respectively,
$\eta_{nk}^r$ and $\eta_{nk}^i$ are the real and imaginary parts
of the additive noise in data.

It is important to built a noise-rejection mechanism into restoring
algorithm. This results in further regularizing  the outputs over and above
what maximizing entropy can do. Assume the presence of the additive circular
Gaussian noise. Let $\sigma_{nk}^2$ be the noise variances for each of
the real and imaginary parts at the spatial frequencies with coordinates
$(n,k)$. Then the net estimation principle becomes

\begin{equation}
 H(x)-1/2 \sum_n~\sum_k (\eta_{nk}^{r2}+\eta_{nk}^{i2})/\sigma_{nk}^2 = \max.
\end{equation}

This is equivalent to the search for a solution $(x, \eta^r, \eta^i)$
that maximizes the joint probability of unknown signal $x_{ml}$ and
noise values $\eta_{nk}^r$ and $\eta_{nk}^i$. By Bayes' theorem with
{\it a priori} probability law for $x$ given by $\exp[H(x)]$ this is
also a maximum {\it a posteriori} probability estimate of the unknowns.

\subsection{The generalized maximum entropy method for estimation of
negative/positive signals}

Formulation (1) of maximum entropy can be applied exclusively to
positive-definite real signals. However we want to work with the Shannon
entropy of the real CMB fluctuations taking both positive and negative values.
Such a real bipolar signal $x_{ml}$ can be represented as the difference
between two positive-definite functions (Bajkova\cite{1}):

\begin{equation}
x_{ml}=x_{ml}^{+}-x_{ml}^{-}, ~~~~~~~x_{ml}^{+},~~x_{ml}^{-} \ge 0.
\end{equation}

Signals $x_{ml}^+$ and $x_{ml}^-$ do not have the overlapping support regions.
Let us replace the problem of estimation of $x_{ml}$ by that of estimation
of $x_{ml}^+$ and $x_{ml}^-$:

\begin{equation}
-\sum_m\sum_l (x_{ml}^{+}\ln x_{ml}^{+}+x_{ml}^{-}\ln x_{ml}^{-})\equiv H(x).
\end{equation}

The nonoverlapping support requirement for $x_{ml}^+$ and $x_{ml}^-$ would not
necessarily be met simply by maximizing (6). As shown in Bajkova\cite{1}, a
tuning parameter, $\alpha$, must be inserted to form the associated entropy
$$
H (x,\alpha)\equiv
$$
\begin{equation}
-\sum_m\sum_l (x_{ml}^{+}\ln(\alpha x_{ml}^{+})+x_{ml}^{-}\ln (\alpha x_{ml}^{-})).
\end{equation}

The value of $\alpha$ is at the discretion of the user. The action of $\alpha$
is to force nonoverlap as $\alpha$ is made larger.

\WFigure{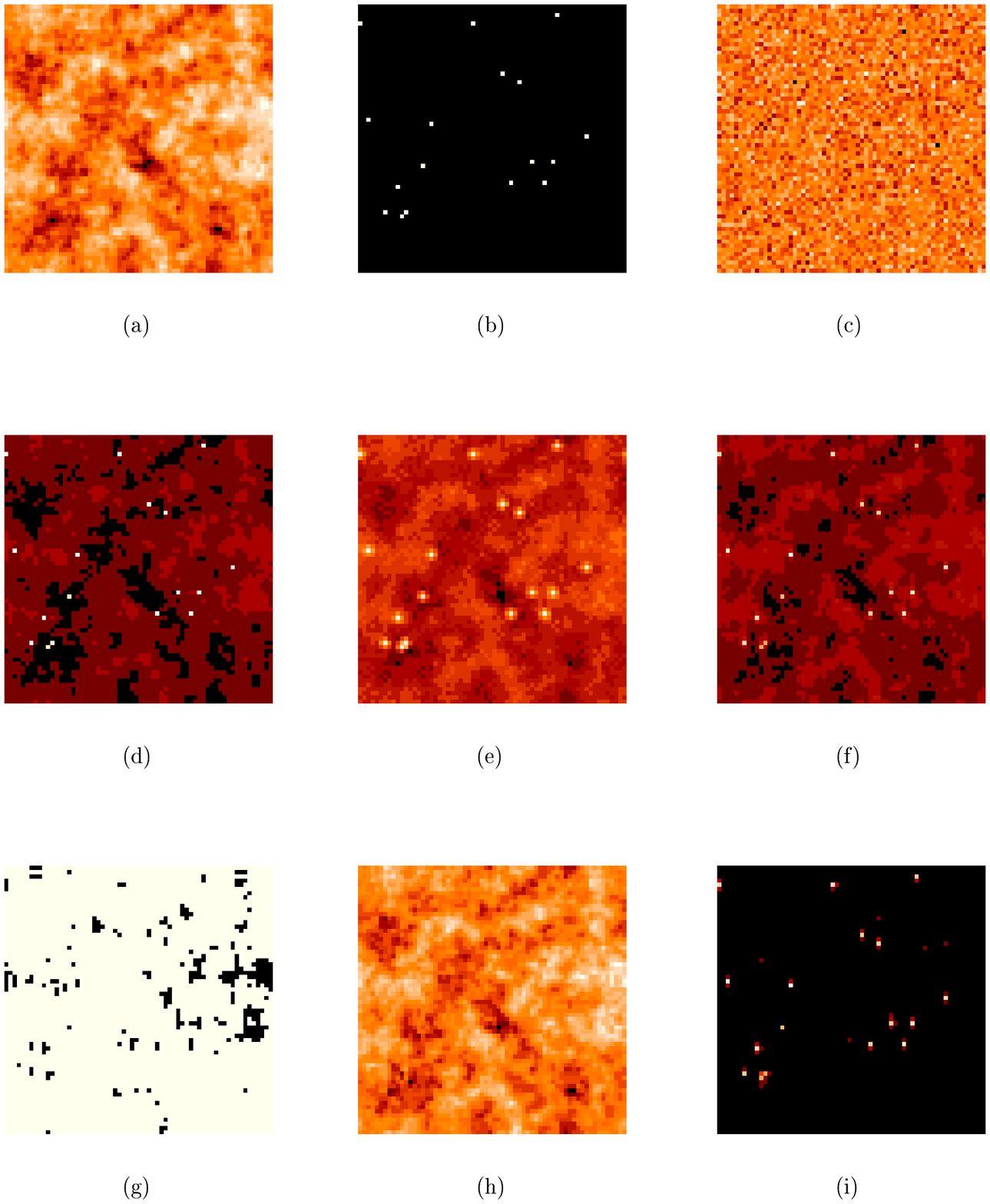}
       {Simulated maps: initial $CMB$ (a); $PS$ (b); initial noise $N$ (c);
       $CMB+PS$ (d); $(CMB+PS)*W + N$ (e); $(CMB+PS)_{rec}$ map (f);
       coordinates of the CMB discarded components (g);
       $CMB_{rec}$ (h); $PS_{rec}$(i)}

\section{Model of the CMB detected map}

For simplicity only the small-angle patch of the sky is considered. Then the CMB
fluctuations $\Delta T = T- <T>$ may be generated by evaluating the simple
Fourier series (Bond et al\cite{2}):

$$
\frac{\Delta T(\theta_x, \theta_y)}{T} =
$$
\begin{equation}
=\sum_{n_u=0}^{N_u-1}\sum_{n_v=0}^{N_v-1} D(n_u, n_v)\exp[i \frac{2\pi}{L}(n_u\theta_x + n_v\theta_y)],
\end{equation}
where $L$ denotes the size of the simulated region;
$(\theta_x, \theta_y)$ are Cartesian coordinates on the sky (spatial domain);
$(n_u,n_v)$ are coordinates of Fourier components $D$ in spatial frequency
domain. For Gaussian CMB fields the amplitude of $D(n_u,n_v)$ satisfies
a Gaussian distribution with zero mean and variance
\begin{equation}
<|D(n_u,n_v)|^2> = C(l),~~l=\frac{2\pi}{L}\sqrt{n_u^2+n_v^2}
\end{equation}
and phases at random in the interval $(0, 2\pi)$, where $C(l)$ is the power
spectrum associated with a spherical harmonic expansion of the radiation
temperature, $l$ is the multipole number (Bond et al\cite{2}).
The PS  background is considered as a point-set of $\delta-$functions randomly
distributed over the sky patch.
In our model the CMB detected (input) map can be represented as

\begin{equation}
CMB_{det}~=~(CMB~+~PS)~*~W~+~N,
\end{equation}
where $*$ denotes the operation of convolution, $W$ is a beam profile, $N$ is
a pixel noise.

\section{The CMB component separation algorithm}

First let us consider the noise-free case when

$$
CMB_{det}~=~(CMB~+~PS)~*~W.
$$

For separation of the primary CMB and PS foreground we propose the following
algorithm.

1. Deconvolution of the CMB input map by the antenna beam profile to obtain the
estimation of the CMB+PS map being denoted as $(CMB+PS)_{rec}$:

$$
(CMB+PS)_{rec} = CMB_{det}~*~ W^{-1}.
$$

In doing it we maximize the generalized entropy functional in accordance with (7)
because the sought for solution for $(CMB+PS)_{rec}$ can take both positive and
negative values. As input data we use all the Fourier components of the detected
CMB input map. The linear constraints (2) and (3) in this case should be
rewritten as

$$
\sum_m\sum_l x_{ml}\cos(2\pi(nm+kl)/M) w^{nk}+\eta_{nk}^r=D_{nk}^r,
$$

$$
\sum_m\sum_l x_{ml}\sin(2\pi(nm+kl)/M) w^{nk}+\eta_{nk}^i=D_{nk}^i,
$$
where ${w^{nk}}, n,k=0,...,M-1$ is the antenna beam function in Fourier domain;
$x_{ml}$ must be represented as (5).

2. Reconstruction of the PS foreground using the standard MEM functional (4)
determined for the positive-definite signals, from only those Fourier spectrum
components of the input CMB map, which are located in hi-multipole region
$(l_d,l_f)$.

Because the data used for the reconstruction of the PS foreground contain a noise due
to the presence in them of the primary CMB signal, we expect that the PS
distribution is reconstructed with perceptible errors. Therefore the direct
subtraction of the reconstructed PS foreground from the $(CMB+PS)_{rec}$ map
can provoke more perceptible errors in CMB estimation, than the simple
discarding from the $(CMB+PS)_{rec}$ map the pixels, contaminated by PS (as
proposed by Tegmark et al\cite{3}). Therefore to increase the precision of the
primary CMB reconstruction we propose the implementation of the following two steps.

3. Discarding from the $(CMB+PS)_{rec}$ map the pixels with coordinates of those
PS reconstructed in the previous step, which amplitude is greater than some
value of the threshold. This operation produces the estimation of primary CMB  with
missing components. The threshold value may be chosen to be comparatively small
in order to ensure the discarding from the CMB map as many as possible number
of the contaminated pixels. But this number should not be too large in order
to ensure the acceptable MEM reconstruction in the next step of the algorithm.

4. Reconstruction of the missing CMB components using a modified GMEM technique
which principally takes into account {\it a priori} information about the most
general features of CMB signal such as finiteness and spherical symmetry of
its power spectrum. (Note that finiteness in spectrum domain means analyticity
of signal in space domain.)

In the case of pixel-noise data the CMB detected map satisfies to equation (10).
To take into account the pixel noise we propose to introduce new unknowns
for the Gaussian pixel noise to include additional terms into the entropic
functional and data constraints in parallel with the unknown signal.
Then the entropic functional to be maximized, represented in the most general form
considering that both signal and noise are bipolar, looks like :

$$
H(x,\alpha)\equiv-\sum_m\sum_l x_{ml}^{+}\ln(\alpha x_{ml}^{+})+x_{ml}^{-}\ln(\alpha x_{ml}^{-})-
$$
\begin{equation}
-\sum_{m}\sum_{l}N_{ml}^{+}\ln(\alpha N_{ml}^{+})+N_{ml}^{-}\ln(\alpha N_{ml}^{-})=\max
\end{equation}

\noindent and the constraints provided by the data are as follows:

$$
\sum_m\sum_l(x_{ml}^{+}-x_{ml}^{-})a_{ml}^{nk}w^{nk}+
$$
\begin{equation}
+\sum_m~\sum_l(N_{ml}^{+}-N_{ml}^{-})a_{ml}^{nk}=D_{nk}^r,
\end{equation}

$$
\sum_m\sum_l(x_{ml}^{+}-x_{ml}^{-})b_{ml}^{nk}w^{nk}+
$$
\begin{equation}
+\sum_m~\sum_l(N_{ml}^{+}-N_{ml}^{-})b_{ml}^{nk}=D_{nk}^i.
\end{equation}

It is the only difference from the separation algorithm described in the previous
subsection.

\section{Simulation results}

To test our approach we have simulated a large number of experiments for both
noiseless and noisy CMB observational data including the PS foregrounds of
different kind. All the tests for noiseless data have given approximately
exact solutions for the CMB separation. The experiment demonstrated here has
been aimed to separate the primary CMB and PS components in presence of
significant pixel noise. For generation of the 7.5 degree square realization of
the primary CMB fluctuations we used the CMB power spectrum function satisfying
the $\Lambda$CDM model with parameters ($\Omega_b h^2=0.02,
\Omega_{\Lambda}=0.65, \Omega_m=0.3, h=0.65, n=1$). The CMB map has been
numerically generated as the Gaussian field in accordance with equations (8)
and (9) mentioned in section 3. The size of discrete maps has been chosen as
64x64 pixels, so that the spatial spectrum of the CMB map is extended up to
$l=1536$. The PS generation, reflecting the most usual case, consists of a
number comparatively bright point-sources randomly distributed over the chosen
sky patch. The pixel-noise map has been generated as the Gaussian field with
zero mean and variance approximately equal to variance of the primary CMB.
Antenna beam profile W has been constructed as the circular symmetric Gaussian
function with a FWHM of 3 pixels.

To realize the deconvolution operation we used the GMEM modification outlined
in section 4. Simulation results are exposed in Figure 1, where (a) is the true
primary CMB generation, (b) is the true PS foreground, (c) is the  generation of
the additive Gaussian pixel noise, (d) is the true CMB+PS map, (e) is the
$CMB_{det}$ map satisfying the equation (10), (f) is $(CMB+PS)_{rec}$ map
obtained by use of the modified GMEM in accordance with equations (11),(12),
(13), (g) is the distribution of the discarded contaminated pixels,
determined by a chosen threshold value, (h) is the $CMB_{rec}$ map obtained by
the GMEM interpolation algorithm using {\it a priori} CMB power spectrum
information, (i) is the reconstructed PS foreground obtained by subtraction
from the $(CMB+PS)_{rec}$ map of the $CMB_{rec}$ map. The corresponding
power angular spectra are shown in Figure 2. The variances of all the
generated, reconstructed and residuals maps are listed in Table~1 .

\Figure{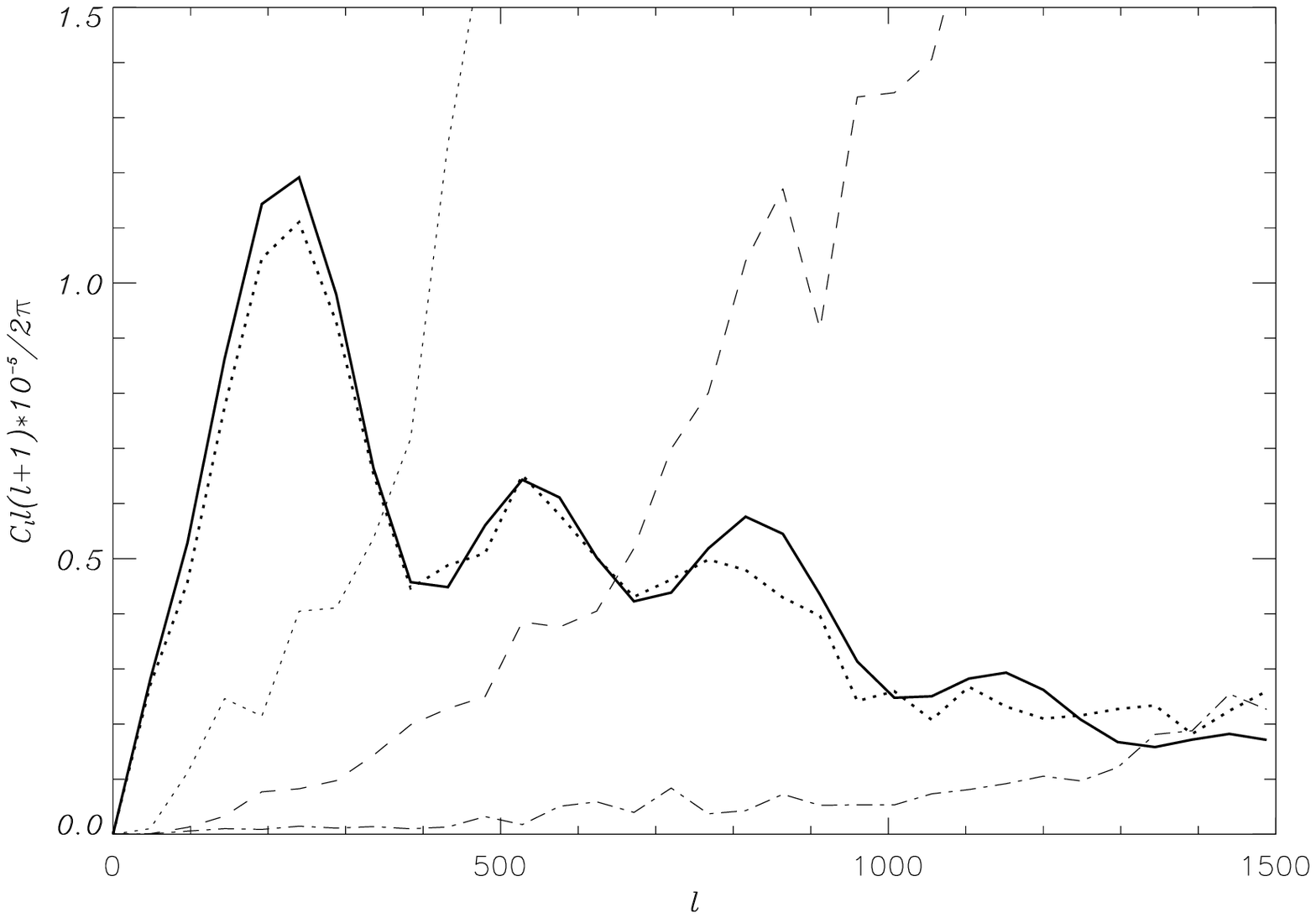}
      {Power angular spectra of:
      initial CMB map (solid), initial PS map (thin dotted);
      pixel noise map (dashed);
      CMB reconstructed map 2) (bold dotted);
      corresponding residual map (dash-dotted)}

\begin{table}
\begin{small}
\caption{Signals and their variances}
\begin{center}
\begin{tabular}{llllll}
\hline
\noalign{\vskip 1mm}
Map & Variance~~~~~~~~ \\
\noalign{\vskip 1mm}
\hline
\noalign{\vskip 1mm}
$CMB$ initial & 390.156 \\
$PS$ initial & 1692.944 \\
Pixel Noise & 345.970    \\
$CMB_{rec}$ & 366.420    \\
Residuals   &            \\
$CMB-CMB_{rec}$ & 19.330 \\
\noalign{\vskip 1mm}
\hline
\end{tabular}
\end{center}
\end{small}
\end{table}

\section{Conclusions}

To summarize our results, we have proposed and successfully tested some
effective realizations of the maximum entropy method (including a generalized
version for the distributions with both positive and negative values) for the
separation from the detected CMB maps of the hi-multipole foregrounds from
extragalactic point sources even in presence of significant pixel noise.
The whole foreground separation algorithm consists of four steps. The first
one assumes the operation of deconvolution of the detected CMB map by the
antenna beam profile. In the second step we reconstruct the PS foreground by
the standard MEM using as initial data the hi-multipole spectrum region of the
CMB+PS recovered map. In the third step we discard from the CMB+PS map the
pixels contaminated by the reconstructed PS, and in the last step we
reconstruct the CMB discarded components by the GMEM procedure using {\it a
priori} information about finiteness and spherical symmetry of the CMB power
spectrum satisfying the Gaussian statistics.


\small
\begin{thebibliography}{1} \itemsep=-5pt

\bibitem{1}

 A.T. Bajkova, {\it A\&ApTr \/} {\bf 1}, 313 (1992).

\bibitem{2}

 J.R. Bond and G. Efstathiou, {\it  MNRAS \/} {\bf 226}, 655 (1987).

\bibitem{3}

 M. Tegmark and A. de Oliveira-Costa, ~~astro-ph/ 9802123, 1998.

\end{thebibliography}
\end{document}